\documentclass[twocolumn,showpacs,amsmath,amssymb]{revtex4}

\usepackage{graphicx}% Include figure files
\usepackage{dcolumn}% Align table columns on decimal point
\usepackage{bm}% bold math
\def\pf6{(TM\-TSF)$_2$\-PF$_6$}
\def\clo4{(TM\-TSF)$_2$\-ClO$_4$}
\begin{document}

\title{Scaling Behavior of the Longitudinal and Transverse Transport \\
in Quasi One-Dimensional Organic Conductors}

\author{M. Dressel}
\author{K. Petukhov}
\author{B. Salameh}
\author{P. Zornoza}
\affiliation{1.~Physikalisches Institut, Universit{\"a}t Stuttgart,
Pfaffenwaldring 57, D-70550 Stuttgart, Germany}
\author{T. Giamarchi}
\affiliation{DPMC, University of Geneva, 24 quai Ernest-Ansermet,
CH-1211 Geneva, Switzerland}
\date{\today}
\begin{abstract}
We report on dc and microwave experiments of the low-dimensional
organic conductors \pf6\ and \clo4\ along the $a$, $b^{\prime}$,
and $c^*$ directions. In the normal state of \pf6\ below $T=70$~K,
the dc resistivity follows a power-law with $\rho_a$ and
$\rho_{b^{\prime}}$ proportional to $T^2$ while $\rho_{c^*}\propto
T$. Above $T = 100$~K the exponents extracted from the data
for the $a$ and  $c^*$ axes are consistent with what is to be expected for a system of
coupled one-dimensional chains (Luttinger liquid) and a
dimensional crossover at a temperature of about $100$~K. The
$b^\prime$ axis shows anomalous exponents that could be attributed
to a large crossover between these two regimes. The contactless
microwave measurements of single crystals along the
$b^{\prime}$-axis reveal an anomaly between 25 and 55 K which is
not understood yet. The organic superconductor \clo4\ is more a
two-dimensional metal with an anisotropy
$\rho_a/\rho_{b^{\prime}}$ of approximately 2 at all temperatures.
Such a low anisotropy is unexpected in view of the transfer
integrals. Slight indications to one-dimensionality are found in
the temperature dependent transport only above 200~K. Even along
the least conducting $c^*$ direction no region with semiconducting
behavior is revealed up to room temperature.
\end{abstract}


\maketitle

\section{Introduction}
The properties of quasi one-dimensional conductors are of
particular interest from a theoretical point of view, because in
one dimension electron-electron interactions lead to a break-down
of the Fermi-liquid picture established half a century ago by Lev
Landau \cite{Landau57,Pines66}. A description by a
Tomonaga-Luttinger liquid \cite{Tomonaga50,Luttinger63} seems to
be more appropriate implying a number of very distinct features
like spin-charge separation and power-laws in certain quantities
\cite{Voit95,Giamarchi04}. While the theory of
one-dimensional electronic systems was put forward by numerous
contributions
\cite{Emery79,Solyom79,Haldene81,Schulz91,Giamarchi91} during the
last decades, the experimental realization is only tackled since a
few years. Attempts in the field of semiconductor quantum wires,
edge states and stripe phases in the quantum Hall effect, carbon
nanotubes, and rows of metal atoms on vicinal surfaces have been
successful to some degree. In particular the quasi one-dimensional
organic conductors of the Bechgaard family
\cite{Jerome82,Jerome94,Ishiguro98} serve as model systems to test
the theoretical predictions. These compounds are charge transfer
salts consisting of stacks of the planar organic molecules TMTSF
(which stands for tetra\-methyl\-tetra\-selena\-fulvalene) along
the $a$-axis which are separated in $c$-direction by monovalent
anions like PF$_6^-$, AsF$_6^-$, ReO$_4^-$, or ClO$_4^-$.
In $b$-direction the distance of the stacks is comparable to the van der Waals radii.
Most prominent findings are the reduced density of states at the
Fermi energy as indicated by photoemission spectroscopy
\cite{Dardel93}, the $c$-axis transport investigated by pressure
dependent dc resistivity \cite{Moser98}, the scaling behavior in
the optical conductivity \cite{Dressel96}, the Hall effect
\cite{Moser00,Mihaly00}, and finally indications of
spin-charge separation by the similarity in the spin dynamics
\cite{Dumm00} and thermal conductivity \cite{Lorenz02} for TMTSF
and TMTTF salts although the electronic transport is very much
different; for a review see Ref.~[\onlinecite{Dressel03}].

Real materials always have a finite coupling between the chains,
no matter how anisotropic they are. The question arises whether
the Luttinger liquid effects can be observed in quasi
one-dimensional systems. The general expectation is that these
compounds cross over from Luttinger liquid behavior to a coherent
behavior as the temperature or frequency is lowered
\cite{Bourbonnais84,Brazovskii85,Bourbonnais91,Clarke94,Schulz96,Biermann01};
the details, however, are still under debate. For a quasi
one-dimensional system with coupling $t_c\ll t_b \ll t_a$, the
effective dimensionality depends on the energy range of interest:
at low temperatures ($k_BT<t_c$) the system is three-dimensional
and only at elevated temperatures ($k_BT>t_b$) or high frequencies
($\hbar\omega
>t_b$), one-dimensional properties are expected. In the case of the
Bechgaard salts, for instance, the transfer integrals are
approximately $t_a:t_b:t_c=250~{\rm meV}:20~{\rm meV}:1~{\rm
meV}$, so the bare crossover integral should be of the order of
200~K. Although it was initially believed \cite{Bourbonnais84}
based on NMR data that this scale would be renormalized by
interactions down to 20~K, the more recent optical and transport
measurements place \cite{Dressel96,Moser98} this scale at about
100~K.

Only very recently attempts were undertaken to describe a system
of weakly coupled Luttinger chains and actually focus on the
interchain transport
\cite{Lopatin01,Biermann01,Biermann02}. From
an experimental point of view, measurements of quasi
one-dimensional organic samples in the direction perpendicular to
the needle axis are extremely challenging and only very few
results on TMTSF salts have been published
\cite{Moser98,Jacobsen81,Henderson99,Zamorszky99,Fertey99,Korin03,Murata81,Forro84}. Here we
report on temperature dependent dc and contactless microwave
measurements of the electrical conductivity in all three
directions of \pf6\ and \clo4.

\section{Experimental Details}
Single crystals of the Bechgaard salts (TMTSF)$_2X$ are grown by
electrochemical methods in an H-type glass cell between room
temperature and $0^{\circ}$C. A constant voltage of 1.5~V was applied
between platnium electrodes with an area of approximately 3~cm$^2$. The
current through the solution was between 9.2 and 13.4$~\mu$A. To reduce
the diffusion, a sand barrier can be introduced. After several months
we were able to harvest needle-shaped to flake-like single crystals of
several millimeters in length and a considerable width
($b^{\prime}$-direction) up to 2~mm.
Due the triclinic symmetry, $b^{\prime}$ denotes the projection of the
$b$ axis perpendicular to $a$, and $c^*$ is normal to the $ab$ plane.
The dc resistivity of \pf6\ along
the $a$-axis  was measured on needle-shaped samples with a typical
dimension of $2~{\rm mm}\times 0.5~{\rm mm}\times 0.1~{\rm mm}$ along
the $a$, $b^{\prime}$, and $c^*$ axes, respectively. The results on the
$b^{\prime}$-axis conductivity were obtained on a narrow slice cut from
a thick crystal perpendicular to the needle axis; the typical
dimensions of so-made samples were $a~\times~b^{\prime}~\times~c^{*} =
0.2~{\rm mm}\times 1.3~{\rm mm} \times 0.3~{\rm mm}$. Single crystals
of \clo4\ have about the same size in $a$ and $b^{\prime}$ directions
(they grow even wider since they are more two-dimensional), however,
the thickness ($c^*$ axis) rarely exceeds 50 or 60~$\mu$m. Due to our
advances in achieving large sample geometry, we were able to measure
$b^{\prime}$-axis resistivity for the first time with basically no
influence of the $a$ and $c^*$ contributions and using standard
four-probe technique to eliminate the contact resistances. Also for the
$c^*$-axis transport, we were able to apply four contacts, two on each
side of the crystal. The contacts were made by evaporating gold pads on
the crystal, then 25~$\mu$m gold wires were pasted on each pad with a
small amount of silver or carbon paint. The \pf6\ samples were slowly
cooled down to avoid cracks and ensure a thermal equilibrium. In the
case of \clo4\ we conducted experiments in the relaxed state using a
slow cooling rate of less than 0.2~K/min. Employing a $^3$He cryostat,
we cooled down below the superconducting transition at $T_c\approx 1.1$~K.
In order to reach the
quenched state, the crystal were cooled down rapidly from about 50~K
with a rate of more than 50~K/min and the data were subsequently
recorded on warming up.

Besides dc experiments we investigated the anisotropic transport of
\pf6\ and \clo4\ single crystals in all three directions at microwave
frequencies. The major advantage of this method is that no contacts
have to be applied. They always lead to a current injection which is
not well defined; it cannot be avoided that the contact pads required
for dc experiments influence the current flow. It is not always clear
which part of the bulk material actually carries the current and which
part is probed for the voltage drop. Also surface currents may have a
large influence in highly anisotropic conductors. In microwave
experiments the dielectric response of the entire sample is integrated;
admittedly other factors, like the geometry, the depolarization factor
or skin effect cause uncertainties.

 For measuring the microwave
conductivity in $a$-direction we employed naturally grown needles of
typical dimensions of $1~{\rm mm}\times 0.2~{\rm mm} \times 0.2~{\rm
mm}$, in the case of \pf6-crystals, for instance. Since a needle-like
geometry is best also for precise microwave measurements, we again cut
a slice ($a~\times~b^{\prime}~\times~c^{*} = 0.2~{\rm mm}\times
1.2~{\rm mm} \times 0.2~{\rm mm}$) from a thick single crystal to
measure in $b^{\prime}$ direction. In order to perform micro\-wave
experiments along the $c^*$ axis, we carefully cut a crystal into
several pieces (approximately cubes of 0.2~mm corner size) and arranged
up to four as a mosaic in such a way that a needle-shaped sample of
about $0.2~{\rm mm}\times 0.2~{\rm mm} \times 0.8~{\rm mm}$ was
obtained \cite{Petukhov03}. In the case of \clo4\ we had to assemble
eleven thin pieces on top of each other to obtain a mosaic of
$a~\times~b^{\prime}~\times~c^{*} = 0.15~{\rm mm}\times 0.2~{\rm mm}
\times 0.66~{\rm mm}$ \cite{Zornoza04}.

The microwave experiments utilize three different cylindrical copper
cavities which resonate in the TE$_{011}$ mode at 24, 33.5, and 60~GHz.
They are fed by voltage-driven Gunn oscillators via suitable waveguides
and operate in the transmission mode. The coupling is about 10\%\ and
done via two holes in the sidewalls. The crystals are positioned in the
maximum of the electric field placed onto a quartz substrate (0.07~mm
thick) and can be rotated in situ around one axis. The samples were
cooled down slowly (0.2~K per minute, to avoid microcracks) from 300~K
to 2~K by coupling to the liquid helium bath with the help of
low-pressure He exchange gas and by utilizing a regulated heater.
Temperatures as low as 0.7~K could be achieved with a special cavity
attached to a $^3$He chamber \cite{Thoms04}. The
stability is better than 10~mK \cite{Petukhov03}. By recording the
center frequency $f$ and the halfwidth $\Gamma$ of the resonance curve
as a function of temperature and comparing them to the corresponding
parameters of an empty cavity ($f_0$ and $\Gamma_0$), the complex
electrodynamic properties of the sample, like the surface impedance,
the conductivity and the dielectric constant, can be determined via
cavity perturbation theory; further details on microwave measurements
and the data analysis are summarized in \cite{Klein93}. The microwave
experiments on \clo4\ using a fast cooling rate of approximately
12~K/min (in order to prevent the tetrahedral anions to order) could
not fully achieve the quenched state, but in some runs we still see
slight bumps in the resistivity at the ordering temperature of $T_{\rm
AO}=24$~K. Thus we concentrate here on the relaxed state obtained by
cooling with less than 0.2 K per minute; in this case the
spin-density-wave state does not show up below $T_{\rm SDW}\approx 6$~K
\cite{remark2}. Since for 60 GHz no complete set of data is available
and no unexpected fact are found by now, we will disregard these
experiments and focus on 24 and 33.5~GHz.

\section{Results and Analysis}
\subsection{\pf6}
\subsubsection{DC Resistivity}
\begin{figure}
\centerline{\includegraphics[width=8cm]{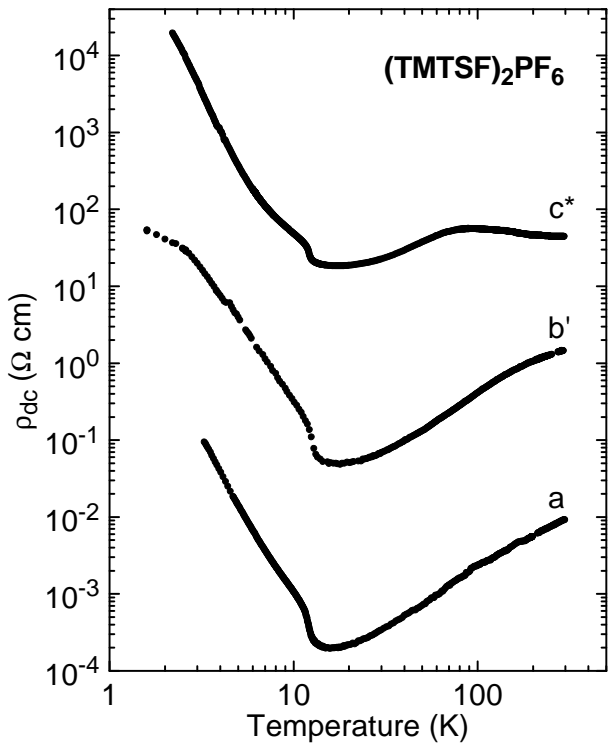}}
\caption{\label{fig:dcpf6total}Overall behavior of the temperature
dependent dc resistivity of a \pf6\ single crystal along the three crystallographic axes
$a$, $b^{\prime}$, and $c^*$.}
\end{figure}
\begin{figure}
\includegraphics[width=8.6cm]{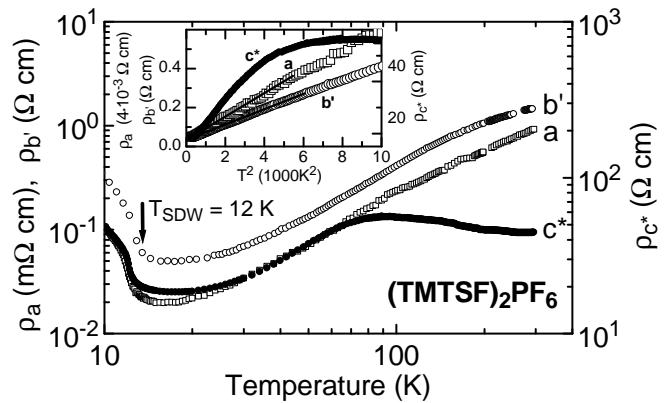}
\caption{\label{fig:dcpf6} Temperature dependence of the dc resistivity
of \pf6\ measured along the $a$ (open squares), $b^{\prime}$ (open
circles), and $c^{\ast}$ (solid circles) directions. Note the
resistivity scales for the three axes. The inset depicts the $T^{2}$
dependence of the resistivity below 100~K. Below 65~K, $\rho_{c^{*}}(T)$
cannot be described by a quadratic dependence; instead it follows a
$\rho_{c^{\ast}} (T)\propto T$ behavior.}
\end{figure}

The temperature dependent dc resistivity for all three directions is
displayed in Fig.~\ref{fig:dcpf6total}. The $a$-axis transport of \pf6\
is metallic above $T_{\rm SDW} = 12$~K with a change in slope around
100~K. At high temperatures it can be fitted to a $\rho_{a}(T) \sim
T^{1.3}$ power-law, while for $T<70$~K the $a$-axis resistivity follows the law
$\rho_{0} + AT^{2}$, as can  be nicely seen in the inset of
Fig.~\ref{fig:dcpf6}. For our samples, values of $\rho_{0} = 1.1 \times
10^{-4}~\Omega$cm and $A = 0.2~\mu \Omega$cmK$^{-2}$ are found; the low
residual resistivity $\rho_{0}$ together with a large resistivity ratio
$\rho_{\rm 300K}/\rho_{\rm 20K}$ indicate a very high crystal quality.
The behavior  agrees well with previously published
data~\cite{Bechgaard80,Jacobsen81,Tomic91,Moser98,Zamorszky99}.

\begin{figure}
\centerline{\includegraphics[width=8.5cm]{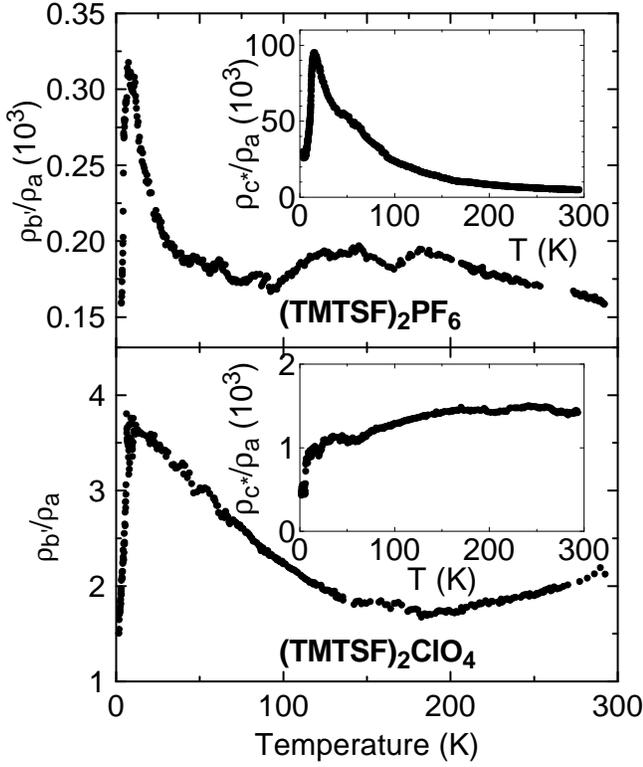}}
\caption{\label{fig:anisotropy}Temperature dependence of the anisotropy
of the dc resistivity $\rho_{b^{\prime}}/\rho_a$ of the Bechgaard salts. The inset
show the temperature dependence of the anisotropy
$\rho_{c^*}/\rho_{a}$. The upper frame corresponds to \pf6\ while the
data of the lower frame are taken on \clo4.}
\end{figure}

To our knowledge, there are only two reports \cite{Jacobsen81,Mihaly00}
on dc measurements of the $b^{\prime}$-axis conductivity of \pf6\
crystals; as a matter of fact both strongly contradict each other. Our
findings are in very good agreement with the recent report of G.
Mih\'{a}ly {\it et al.} \cite{Mihaly00} where the Montgomery method was
employed. As plotted in Fig.~\ref{fig:dcpf6}, the resistivity in the
intermediate $b^{\prime}$ direction shows a metal-like behavior;
$\rho_{b^{\prime}}(T)$ decreases almost as steeply as $\rho_{a}(T)$.
The kink in $\rho_{b^{\prime}}(T)$ shows up at somewhat higher
temperatures: it follows the dependence $\rho _{b^{\prime}}(T) \propto
T^{0.84}$ between 300 and 200~K and changes to the $T^ {1.63}$
power-law upon further cooling down to 60~K. Below that temperature it
can be perfectly described by a quadratic temperature dependence: $\rho
_{b^{\prime}}(T) \propto T^{2}$ as already found for the $a$ axis (see
inset of Fig.~\ref{fig:dcpf6}).
In Fig.~\ref{fig:anisotropy} the ratio of $\rho_{b^{\prime}}/\rho_a$ is
plotted as a function of temperature for both compounds. For \pf6\ the
ratio is basically constant at higher temperatures above 50~K.

For the least conducting direction, $\rho_{c^*}(T)$ increases by about
a factor of 1.5 when going from room temperature down to 90~K
(Fig.~\ref{fig:dcpf6}). Although no clear power-law is found, the
behavior above 100~K may be approximated by $\rho_{c^*}(T)\propto
T^{-0.2}$. Below 90~K, $\rho _{c^*}(T)$ falls rapidly before turning
upwards again below 15~K due to the spin-density-wave transition. In
the temperature range between 35 and 65~K it follows a metallic
behavior with $\rho_{c^*}(T)\propto T$. Our findings in the $c^{*}$
direction are consistent with  previous
results~\cite{Jacobsen81,Moser98,Mihaly00}. The anisotropy ratio
$\rho_{c^*}/\rho_{a}$ (depicted in the inset of
Fig.~\ref{fig:anisotropy}) increases continuously by a factor of 100 as
the temperature is lowered to $T_{\rm SDW}$ and finally reaches almost
$10^5$. Again fluctuation effects are seen above the transition.

In order to be able to directly compare our results $\rho(T)$ recorded
at ambient pressure with the theoretical models for constant-volume
$\rho^{(V)}(T)$ dependence, we converted our experimental data
utilizing a procedure as it was previously suggested for (TMTSF)$_{2}X$
and (TMTTF)$_{2}X$ salts~\cite{Jerome94,Korin03}. The ambient-pressure
unit cell at 16~K was taken as reference unit cell; when the
temperature $T$ increases, a certain pressure $p$ (depending on the
thermal expansion and the compressibility) must be applied (at a given
$T$) in order to restore the reference volume. Since in the metallic
phase $\rho_{a}$ varies by 10{\%} per 1~kbar~\cite{Moser98,Mihaly00}
for all $T$ values above 50~K, the measured resistivity $\rho_{a}$ is
then converted into a constant-volume value $\rho_a^{(V)}$ using the
expression $\rho_a^{(V)} = \rho_a / \left( 1 + p\cdot 0.1{\rm
kbar}^{-1}\right)$ \cite{remark1}. The analogous procedure using
10{\%}kbar$^{-1}$ and the appropriate thermal expansion can be applied
for the perpendicular direction in order to get
$\rho_{b^{\prime}}^{(V)}$. It is not clear whether this is also a
legitimate procedure with respect to the $c^*$ axis for which a
different transport mechanism applies and the variation with pressure
is not the same for all temperatures. Hence we restrained ourselves
from applying a similar transformation to the $c^*$-axis data without
knowing the exact results.
\begin{figure}
\includegraphics[width=8.6cm]{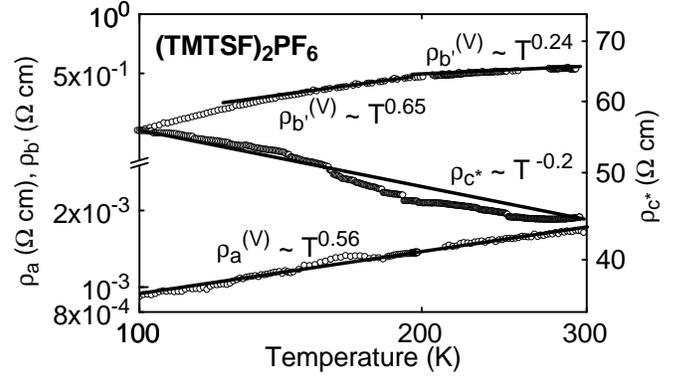}
\caption{\label{fig:dcpf6V} Log-log presentation of the temperature
dependent dc resistivity of \pf6\ in $c^*$ direction: $\rho_{c^*}(T)$
refers to the right axis. Along the $a$ and $b^{\prime}$ direction the
resistivity normalized to constant volume, $\rho_a^{(V)}(T)$ and
$\rho_{b^{\prime}}^{(V)}(T)$, is plotted corresponding to the left
axis. The straight lines indicate to the power-laws.}
\end{figure}

At low temperature, $T <$ 50 K, both the thermal expansion and the
pressure coefficient are small~\cite{Jerome94,Gallois87}.
Therefore, the constant-volume temperature dependence of the
resistivity does not deviate from the quadratic law observed under
constant ambient pressure. In general, the constant-pressure to
constant-volume corrections has the consequence that the
temperature behavior of the dc resistivity yields reduced
power-laws, as can be seen from Fig.~\ref{fig:dcpf6V}. Along the
chain axis, the constant-volume resistivity follows the power-law
$\rho_{a}^{(V)}(T)\propto T^{0.56}$ from room temperature down to
100~K and lower. The temperature dependence of the transverse
$b^{\prime}$-axis constant-volume resistivity does not follow a
single power-law. Nevertheless, the slope may be approximated by
$\rho_{b^{\prime}}^{(V)}(T)\propto T^{0.24}$ and by
$\rho_{b^{\prime}}^{(V)}(T)\propto T^{0.65}$ in the temperature
ranges $200~{\rm K}< T <300$~K and $150~{\rm K}<T<200$~K,
respectively.

\subsubsection{Microwave Experiments}
\begin{figure}
\centerline{\includegraphics[width=7.5cm]{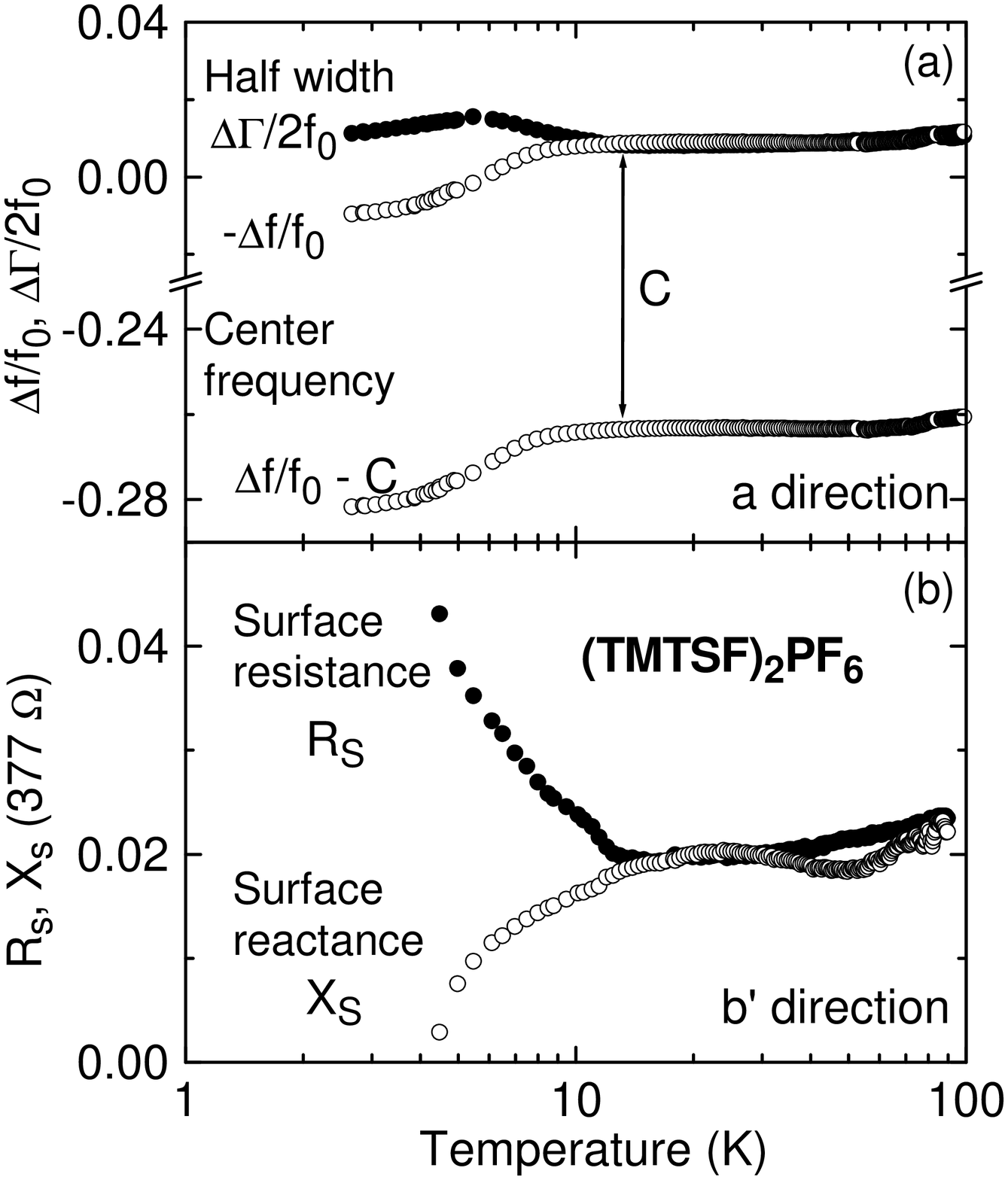}}
\centerline{\includegraphics[width=7.5cm]{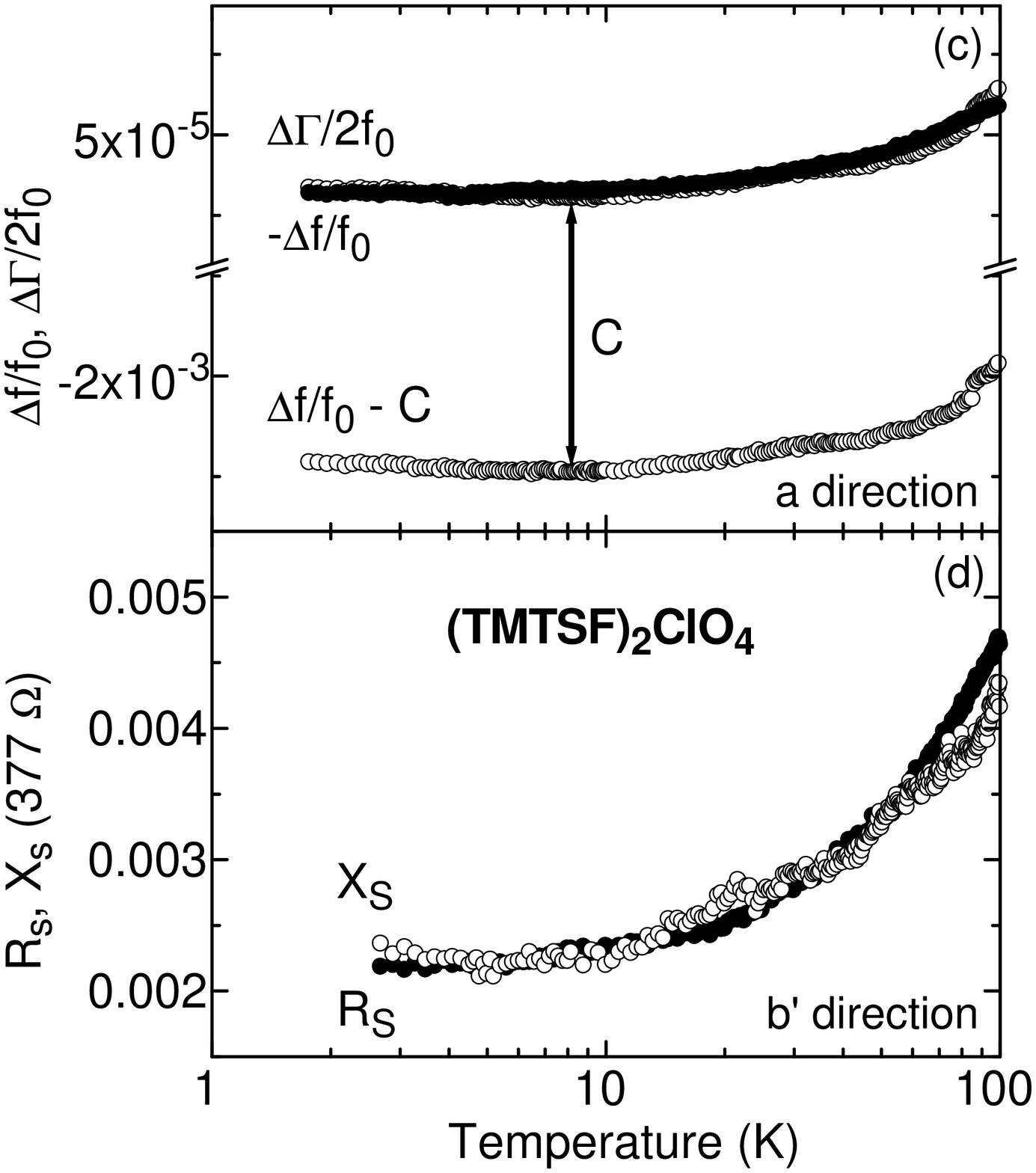}}
\caption{\label{fig:RsXs}(a) The shift in center frequency $-\Delta
f/f_0$ and the halfwidth $\Delta \Gamma /2f_0$ temperature profiles of
a microwave cavity measurement performed at 33~GHz on a \pf6\ single
crystal along the $a$-axis. By adding a constant value $C$ to the
frequency shift $-\Delta f/f_0$ both curves fall on top of each other
over a broad temperature range 12~K$<T<$150~K. This implies that the
Hagen-Rubens approximation applies at these temperatures: $R_S=\Delta
\Gamma/2f_0 \simeq X_S = -\Delta f/ f_0$. (b)~Also in the $b^{\prime}$
direction, the surface resistance $R_S$ and the reactance $X_S$ show
the same temperature behavior, as expected for a metal in the
Hagen-Rubens limit. (c)~Accordingly the temperature dependent shift of
$-\Delta f/f_0$ and $\Delta \Gamma /2f_0$ of \clo4\ along the $a$
direction, and (d)~the temperature dependence of the $R_S$ and $X_S$
along the $b^{\prime}$ direction of \clo4\ measured at 24 GHz. Since
the crystals remain metallic the Hagen-Rubens condition is always
fulfilled.}
\end{figure}
Since \pf6\ is highly conducting along the $a$ direction, the data
analysis is done in terms of the surface resistance $R_S$ and the
surface reactance $X_S$, assuming that the skin-depth is much smaller
than the sample dimension \cite{Klein93}. Our measurements are
performed at microwave frequencies for which we expect the Hagen-Rubens
limit ($\omega \tau \ll 1$) to be appropriate for our analysis; this
was also the case in previous investigations~\cite{Donovan92}. Indeed,
the temperature dependence of the relative change of the halfwidth
$\Delta \Gamma /2f_0=R_S$ and the center frequency $-\Delta f
/f_0=X_S+C$ have the same profile over the wide temperature range, as
depicted in Fig.~\ref{fig:RsXs}a. This is a strong proof that along the
$a$-direction the organic conductor (TMTSF)$_2$PF$_6$ is in the
Hagen-Rubens limit at microwave frequencies, and the surface resistance
$R_S$ and the surface reactance $X_S$ (given in units of the free space
impedance $Z_0=4\pi/c=377~\Omega$) have equal absolute values within an
additive constant $C$ introduced when the cavity is disassembled in
order to mount the sample. Below 12~K the metallic behavior vanishes
since \pf6\ enters the spin-density-wave state. From the surface
impedance the calculation of the complex conductivity
$\sigma_1+i\sigma_2$ is straight forward \cite{DresselGruner02}:
\begin{figure}
\centerline{\includegraphics[width=8cm]{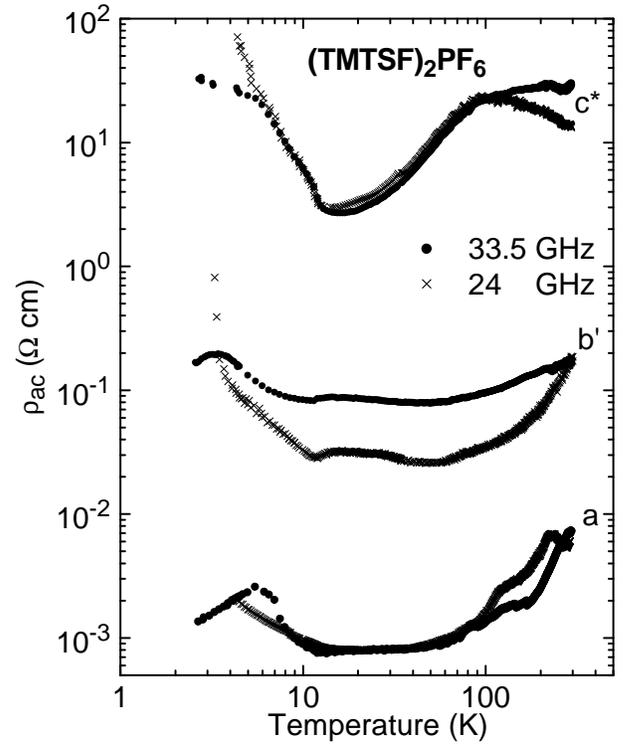}}
\caption{\label{fig:mwpf6total}Temperature dependence of the microwave
resistivity of (TMTSF)$_2$PF$_6$ along the $a$, $b^{\prime}$, and $c^*$
crystallographic axes, measured at 24 and 33.5~GHz.}
\end{figure}
\begin{eqnarray}
\sigma _1 &=& \frac{{f_0 R_S X_S }}{{(R_S^2  + X_S^2 )^2 }},\\
\sigma _2 &=& \frac{{f_0 \left( {X_S^2  - R_S^2 } \right)}}
{{2\left({X_S^2  + R_S^2 } \right)^2 }}.
\end{eqnarray}
The results obtained for \pf6\ at 33.5~GHz are plotted in
Fig.~\ref{fig:mwpf6total}. Most obvious, the resistivity ratio
$\rho_{\rm 300K}/\rho_{\rm 20K}$ is smaller by a factor of 5 in the
microwave range compared to the dc results; this observation was
confirmed by a number of different crystals.

\begin{figure}
\centerline{\includegraphics[width=8cm]{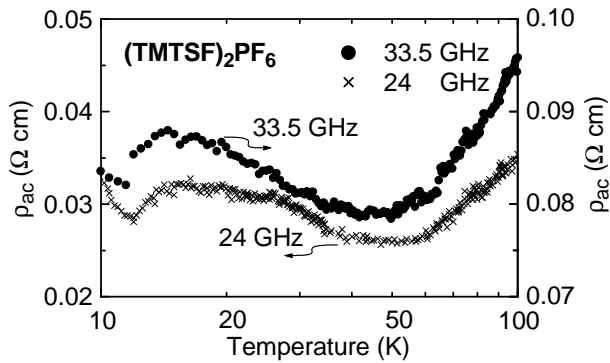}}
\caption{\label{fig:mwpf6b}Enlargement of the temperature dependent
microwave resistivity of \pf6\ along the $b^{\prime}$ axis. }
\end{figure}

For the analysis of our data measured along the $b^{\prime}$ direction,
i.e.\ perpendicular to the stacks, we use the same arguments as for the
$a$ direction. From the dc conductivity measurements on \pf6\ we have
found that $\rho_{b^{\prime}}$ is in the order of 0.1--1~$\Omega$cm,
which gives us a value for skin depth of 0.1~mm at 33.5~GHz. Thus we
expect the system to be in the skin-depth regime along the $b^{\prime}$
direction, at least in the normal state. Again we find that a good
accordance of the surface resistance and surface reactance, as depicted
in Fig.~\ref{fig:RsXs}b. The temperature behavior of the
$b^{\prime}$-axis con\-duc\-ti\-vi\-ty of \pf6\ probed at 33.5~GHz
differs from the dc conductivity along this axis. As can be seen at the
enlargement plotted in Fig.~\ref{fig:mwpf6b}, the profile of the
microwave transport exhibits a change from the metallic behavior to a
semiconductor-like behavior for $T<50$~K. Below a minimum at around 15
to 20~K, $\rho_{b^{\prime}}(T)$ becomes metal-like on further cooling
down to $T_{\rm SDW}=12$~K, where the SDW transition evidences. This
unusual behavior is  robust: it has been observed on a large number of
different samples and in general coincides with previous results of
16.5~GHz-conductivity measurements on (TMTSF)$_2$PF$_6$ along the
$b^{\prime}$ direction, which were performed on
mosaics~\cite{Fertey99}. By now it is not clear whether the slight
shift to lower temperatures with increasing frequency is significant or
not. The Sherbrooke group reports some sample dependence as far as the
detailed shape is concerned.

The dc experiments performed along the $c^*$ axis of \pf6\ yield room
temperature values around $\rho_{c^*}=50~\Omega$cm, leading to a skin
depth much larger than the sample size. Therefore the data on frequency
shift and change of linewidth taken along the $c^*$ direction of \pf6\
were analyzed according to the depolarization regime \cite{Klein93}.
The results obtained on the mosaic assembled from two, three and four
cubes coincide perfectly except for the temperature range between 15
and 40~K where the large sample volume with high losses caused an
overload of the cavity and we could only use the data of the two-block
mosaics \cite{Petukhov03}. The overall temperature behavior of the
$c^*$-axis microwave conductivity of (TMTSF)$_2$PF$_6$ shown in
Fig.~\ref{fig:mwpf6total} is in good agreement with the dc conductivity
data along this direction, except for the temperature region above
80~K, where a slightly semiconducting behavior was observed in the dc
results, while the 33~GHz conductivity is almost temperature
independent in this temperature region.

Following the measurement procedure and data analysis described
above for our 33.5~GHz experiments, we conducted microwave cavity
perturbation measurements also for 24~GHz. The results plotted in
Fig.~\ref{fig:mwpf6total} confirm our findings in any regard;
which is not surprising, since the measurement frequencies are too
close to expect a significant frequency dependence. The only
point to be noticed is the slightly reduced ratio $\rho(15{\rm K})/\rho(300{\rm K})$
along the $b^{\prime}$ direction when going to 33.5~GHz.

\subsection{\clo4}
\subsubsection{DC Resistivity}
\begin{figure}[b]
\centerline{\includegraphics[width=8cm]{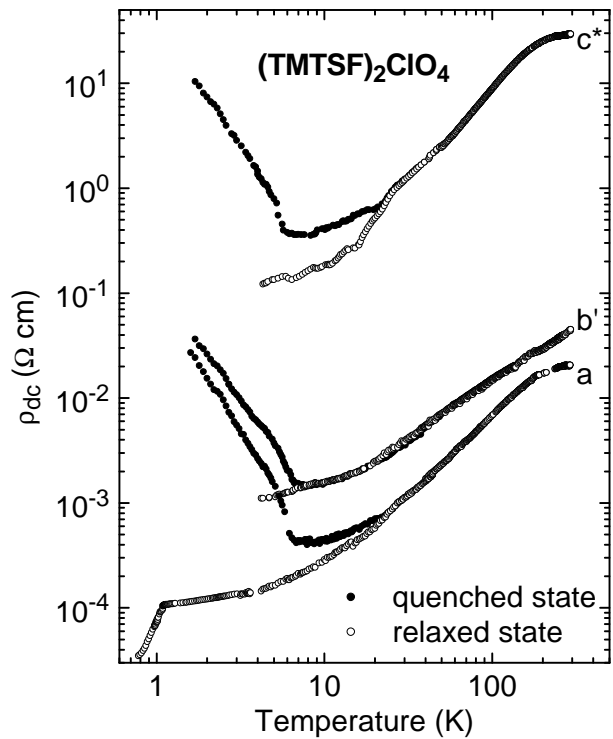}}
\caption{\label{fig:dcclo4total}The dc resistivity of a \clo4\ single crystal along the
three crystallographic axes  $a$, $b^{\prime}$, and $c^*$. The solid
points are taken after a rapid cooling faster than 50~K per minute;
around 6~K, the crystal enters the spin-density wave state. The open
dots correspond to the slow cooled phase which remains metallic and becomes
superconducting at $T_c=1.1$~K.}
\end{figure}
In Fig.~\ref{fig:dcclo4total} the temperature dependence of the dc
transport of \clo4\ is plotted for all three directions in the relaxed
and the quenched state; no results of all three orientations
have previously been published by other groups.
K. Murata {\it et al.} \cite{Murata81} give a room-temperature
anisotropy ratio of $\rho_a:\rho_{b^{\prime}}:\rho_{c^*}\approx 1:23:900$;
their absolute values are in agreement with our findings.
A temperature dependence of the $a$ and $c^*$ direction similar
to our results was reported by L. Forr{\'o} {\it et al.} \cite{Forro84}.
Surprisingly we observe a very low anisotropy $\rho_{b^{\prime}}/\rho_{a}$,
even at room temperature (Fig.~\ref{fig:anisotropy}).
Depending on the cooling rate, a clear difference in the
low-temperature resistivity $\rho_a(T)$, $\rho_{b^{\prime}}(T)$ and
$\rho_{c^*}(T)$ is found below the anion ordering temperature $T_{\rm
AO}=24$~K. If cooled down slowly with a rate of approximately
0.2~K/min, the relaxed state is reached for which the metallic
conductivity continues until the superconducting state is entered at
$T_c=1.1$~K \cite{Thoms04}. In all three directions, a significant
change in slope $\rho(T)$ can be detected around $T_{\rm AO}$, with
only little change in the $b^{\prime}$ direction. Interestingly, while
the resistivity ratio $\rho_{c^*}/\rho_{a}$ increased strongly by a
factor of 50 with decreasing temperature for \pf6\
(Fig.~\ref{fig:anisotropy}), in the case of \clo4\ the anisotropy
remains constant down to about 150~K and becomes smaller at lower
temperatures.

\begin{figure}
\includegraphics[width=8cm]{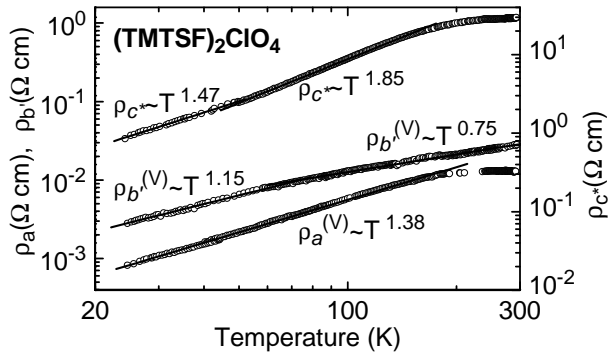}
\caption{\label{fig:dcclo4V} Temperature dependence of the dc
resistivity of \clo4\ above the anion ordering temperature $T_{\rm AO}$
in a double logarithmic representation. $\rho_{c^*}(T)$ is the ambient
pressure resistivity in $c^*$ direction;  along the $a$ and
$b^{\prime}$ direction the data are normalized to constant volume:
$\rho_a^{(V)}(T)$ and $\rho_{b^{\prime}}^{(V)}(T)$. The straight lines
correspond to the power-laws as indicated.}
\end{figure}
Along the chains, the resistivity follows a $\rho_a(T)\propto
T^{1.5}$ law in the temperature range $25~{\rm K}<T<180$~K; below
15~K a linear and even sub-linear behavior is found in the relaxed
state. Above 200~K the resistivity increases more slowly with
temperature, following approximately $T^{0.43}$. The transverse
resistivity $\rho_{b^{\prime}}(T)$ depends linearly on temperature
above 70~K; in the intermediate range ($20~{\rm K}<T<70$~K) a
$T^{1.25}$ behavior is observed; the low-temperature resistivity
($T<10$~K) can be approximated by $\rho_{b^{\prime}}(T)\propto
T^{0.4}$. A similar slope of $\rho_{c^{*}}(T)\propto T^{0.5}$
observed for the least-conducting direction; above $T_{\rm AO}$ up
to 50~K, the resistivity follows a $T^{1.47}$ power-law, which
increases to $T^{1.85}$ for $50~{\rm K}<T<150$~K. Similar to the
$a$ direction, above 200~K $\rho_{c^{*}}(T)$ exhibits a very slow
temperature dependence of $T^{0.34}$.

Following the reasoning given above, we tried to transform the
ambient-pressure results to values of the temperature dependent
resistivity at constant volume: $\rho_a^{(V)}(T)$,
$\rho_{b^{\prime}}^{(V)}(T)$ and $\rho_{c^*}(T)$ are plotted in
Fig.~\ref{fig:dcclo4V}. This attempt is hampered by the lack of
pressure dependent data on the lattice parameter and the resistivity
for \clo4; thus we had to go back and use the transformation procedure
applied for \pf6. Along the chains we find $\rho_a^{(V)}(T)\propto
T^{1.38}$ from the anion ordering temperature $T_{\rm AO}=24$~K up to
almost 200~K; the behavior is temperature independent above. The
$b^{\prime}$-axis response reveals a change in slope around  75~K for
lower temperatures $\rho_{b^{\prime}}^{(V)}(T)\propto T^{1.15}$ while
above the power-law decreases to $T^{0.75}$.

\subsubsection{Microwave Experiments}
As far as we know, this is the first
microwave investigation performed on \clo4\ in all three
crystallographic directions.
The data along the $b^{\prime}$ axis shown in \cite{Henderson99}
are based on the room-temperature anisotropy reported by K.
Murata {\it et al.} \cite{Murata81} and the Hagen-Rubens assumption.
The analysis of our microwave data
obtained on \clo4\ by cavity perturbation technique follows the
procedure described above for our measurements on \pf6. Along the
$a$ and $b^{\prime}$ axes the skin depth is much smaller than the
sample size due to the high conductivity and thus the analysis is
done via the surface impedance (Figs.~\ref{fig:RsXs}c and d). Along
the least conducting $c^*$ axis, the quasi-static depolarization
regime applies \cite{Klein93}.

\begin{figure}
\centerline{\includegraphics[width=8cm]{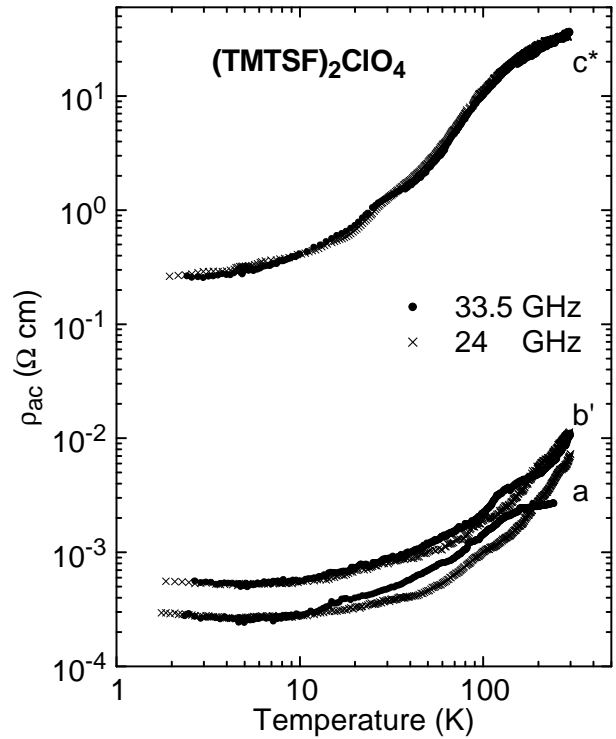}}
\caption{\label{fig:mwclo4total}Temperature dependent microwave
resistivity of \clo4\ along the $a$, $b^{\prime}$, and $c^*$
crystallographic axes, measured at 24 and 33.5~GHz. The data are taken
in the relaxed state.}
\end{figure}

As seen from Fig.~\ref{fig:mwclo4total}, the microwave resistivity
exhibits a metallic temperature dependence in all three orientations,
although the absolute values are very much different.
The temperature dependence of the anisotropy ratios
$\rho_{b^{\prime}}/\rho_a$ and $\rho_{c^*}/\rho_{a}$ obtained
at microwave frequencies is very similar the dc results plotted in
Fig.~\ref{fig:anisotropy}.
Between the $a$ and $b^{\prime}$ axes the anisotropy is basically
temperature independent and approximately $\rho_{b^{\prime}}/\rho_a=2$.
The ratio to the $c^*$-direction is more than three orders of magnitude and
decreases linearly by a factor of 10 when the temperature decreases
below 150~K. Thus \clo4\ resembles a two-dimensional metal much more
than a one-dimensional. These findings are supported by
optical studies on \clo4\ which reveal that the spectral weight of
the zero-energy mode is almost isotropic for the $a$-$b$ plane,
while for the $c^*$ direction no Drude contribution
was revealed \cite{Henderson99}.

Along the $a$ and $b^{\prime}$ axes some of the samples exhibit a
$T^2$ behavior all the way down to 30~K, with no kink observed as
known from \pf6. The $c^*$ resistivity follows a
$\rho_{c^*}(T)\propto T^2$ behavior for $T> 30$~K with decreasing
slope above 130~K. In no temperature regime a semiconducting
behavior is found. The anion ordering at $T_{\rm AO}=24$~K leads
to a slight drop in resistivity with decreasing temperature. For
the $a$ and $b^{\prime}$ directions the resistivity in the range of
lower temperatures is close to linear, but does not follow a
power-law convincingly.

\section{Discussion}

Our results give clear evidence that neither \pf6\ nor \clo4\ can
simply be labelled  a one-dimensional metal. The coupling in the
$b^{\prime}$ direction between the chains has to be considered.
From a theoretical point of view, the transport properties of
coupled one-dimensional chains have been investigated
\cite{Giamarchi91,Lopatin01,Biermann01}. Due to the interactions
the effective interchain hopping is
renormalized \cite{Bourbonnais84}, leading to a smaller crossover
scale than that for free electrons. A simple expression for this
crossover scale is $E^*\propto
t\left(t_{\perp}/t\right)^{1/(1-\alpha)}=
t_{\perp}\left(t_{\perp}/t\right)^{\alpha/(1-\alpha)}$ where
$\alpha$ is the single-particle exponent. For commensurate chains
a more complex analysis is needed to determine $E^*$ (see e.g.\
Ref.~[\onlinecite{Biermann02}]).

The conductivity parallel \cite{Giamarchi91} to the chains
$\sigma_{\parallel}$ and the conductivity perpendicular
\cite{Lopatin01,Biermann01} to the chains $\sigma_{\perp}$ were
calculated in a system of weakly coupled Luttinger chains. It was
found that the interchain hopping is responsible for the metallic
character of the (TMTSF)$_2X$ compounds, which would be otherwise
Mott insulators. The temperature-dependent transport yields a
power-law for the longitudinal and transverse resistivity,
respectively:
\begin{equation}
\rho_{\parallel}(T)\propto (g_{1/4})^2T^{16K_{\rho}-3} \label{eq:Luttinger1}
\end{equation}
\begin{equation} \label{eq:Luttinger2}
\rho_{\perp}(T)\propto T^{1-2\alpha}
\end{equation}
where $g_{1/4}$ is the coupling constant for the umklapp-scattering
process with $1/4$ filling, $K_{\rho}$ is the Luttinger-liquid exponent
controlling the decay of all correlation functions ($K_{\rho}=1$
corresponds to non-interacting electrons and $K_{\rho}<0.25$ is the
condition upon which the 1/4 filled umklapp process becomes relevant),
and $\alpha=1/4\left(K_{\rho}+1/K_{\rho}\right)-1/2$ is the
Fermi-surface exponent.  For the frequency-dependent transport the
conductivity parallel and perpendicular to the chains is given by power-laws:
\begin{equation}
\sigma_{\parallel}(\omega)\propto \omega^{16K_{\rho}-5} \label{eq:Luttinger3}
\end{equation}
\begin{equation}
\sigma_{\perp}(\omega)\propto \omega^{2\alpha-1} \label{eq:Luttinger4}
\end{equation}
Optical experiments on \pf6\ and \clo4\ along the chains
\cite{Dressel96} yield $K_{\rho}=0.23$ in both cases. We now want
to see whether we can describe the temperature dependent
longitudinal and transverse transport in both compounds
consistently. In that respect \pf6\ and \clo4 have to be
distinguished.

\subsection{\pf6}
Indeed \pf6 shows a quite consistent behavior with the above
theoretical description, with a crossover scale of about $E^* \approx
100$~K. Although \pf6\ is highly anisotropic as far as the
absolute values of the resistivity are concerned, the temperature
dependence below $T \approx 100$~K is the same for the $a$ and
$b^{\prime}$ directions implying a similar transport mechanism. In
Fig.~\ref{fig:anisotropy} the ratios of $\rho_{b^{\prime}}/\rho_a$
are plotted as a function of temperature. Above 50~K the ratio for
\pf6\ is basically constant and corresponds to the bandstructure
anisotropy (this does not hold for the $c^*$-direction as we will
discuss below). Using 4$t_{a} = 1.0$~eV together with measured
resistivity gives ($t_{a} : t_{b} : t_{c}) \approx  (250 : 10 :
1)$~meV, which is quite close to the anisotropy ($t_{a} :t_{b} :
t_{c}) = (250 : 17 : 0.75)$~meV as determined by P.~M.~Grant by
band structure calculations~\cite{Grant83}. The increase of
$\rho_{b^{\prime}}/\rho_a$ below 50~K is mainly caused by the
transverse resistivity in $b^{\prime}$ direction which levels off
as the spin-density-wave transition at 12~K is approached. It is
not clear at this point, however, whether fluctuations are solely
responsible for this behavior since they are expected to dominate
along the chains. $\rho_{c^*}/\rho_{a}$ continuously increases by
a factor of 50 when going from room temperature to $T_{\rm SDW}$.
Since the temperature is always larger than the tunnelling
integral along the $c^*$ direction, the conductivity is always
incoherent along this axis and measures in fact the tunnelling
density of states between the planes. Below $70$~K, the dc
resistivity of \pf6\ follows a power-law $\rho_a,
\rho_{b^{\prime}}\propto T^2$, as expected for a Fermi liquid with
electron-electron scattering, and $\rho_{c^*}\propto T$. This last
behavior is consistent with Eq.~(\ref{eq:Luttinger2}) when the planes
are in a Fermi liquid state since then $\alpha = 0$. From the
power-law $\rho_a^{(V)}\propto T^{0.56}$ we found above $T =
100$~K the corresponding Luttinger-liquid exponent can be
calculated to be $K_{\rho} = 0.22$ by using Eq.~(\ref{eq:Luttinger1})
and $\alpha = 0.69$. This value is in good agreement with the
optical data value \cite{Dressel96} quoted above. For the $c^*$
direction we observe $\rho_{c^*}\propto T^{-0.2}$ quite consistent
with Eq.~(\ref{eq:Luttinger2}) and the above value of $K_\rho$ (which
would lead to $T^{-0.29}$). The interpretation of
$\rho_{b^{\prime}}^{(V)}$ is more complex since no simple
power-law was found over the entire temperature range. However a
change in the anisotropy behavior is clearly seen to take place
above 100~K in Fig.~\ref{fig:anisotropy}
and the fit to a power-law in Fig.~\ref{fig:dcpf6V} shows a
downturn of the exponent from 0.65 to 0.24. The data could
thus be interpreted as due to a crossover regime between the
low-temperature Fermi liquid one and the high-temperature Luttinger
liquid. Note that it is reasonable to expect a much larger
crossover region for the $b^{\prime}$-axis transport, than for
the $c^*$ axis given the much higher value of the transfer integral
in this direction.

The optical data along the $c^*$ axis \cite{Henderson99} is compatible with the above
conclusions. The $c^*$-axis resistivity measured at 24~GHz possess
the semiconductor-like behavior (d$\rho_{c^*}/{\rm d} T<0$) in the
high-temperature region $100~{\rm K}<T<200$~K with the power-law
$\rho_{c^*}(T) \propto T^{-0.6}$.  From this power-law we
obtain $K_{\rho}=0.20$, again in reasonable agreement with the
other determinations of the Luttinger-liquid exponent. The
interpretation of the $b^{\prime}$-axis optical data is more
complex. Given the above interpretation of the dc transport the
system is clearly in the two-dimensional regime for temperatures
below $100$~K and still in the crossover regime even up to room
temperature. Thus an interpretation of the optical data along the
$b^{\prime}$ direction can only be done by using a two dimensional theory
for the planes (e.g.\ along the lines of
Ref.~[\onlinecite{Biermann01}]). Note that the temperature
dependence is quite sensitive to the moderate frequency change, as
depicted in Fig.~\ref{fig:mwpf6total}. After performing the
conversion to constant-volume, we obtain
$\rho_{b^{\prime}}^{(V)}(T) \propto T^{-0.4}$ in the temperature
range 20~K$<T<$55~K. This is to be contrasted with the $T^2$
behavior observed in the dc transport in the same temperature
range. The fact that a frequency of an energy corresponding to
$\sim 1$~K gives such a change in the $b^{\prime}$-axis
conductivity signals a very narrow Drude peak, whose behavior
remains clearly to be understood.

\subsection{\clo4}
The situation is more complex for \clo4\ which obviously exhibits
a much more two-dimensional behavior. As can be seen from
Fig.~\ref{fig:dcclo4total}
 the behavior along the $a$ and $b^\prime$ axes is quite
similar over the whole temperature range, while the $c^*$ axis
remains clearly metallic with a flattening only around $300$~K.
This suggests that the crossover temperature in the case of \clo4\
is higher than $200$~K, which means that the dc transport is
totally controlled by the two-dimensional physics. Note that the
optical data \cite{Dressel96} clearly shows for \clo4\ the power-law
behavior of a Luttinger liquid at high energy, which is consistent
with the existence of a crossover scale, but which in that case
would be much higher than for \pf6. However the dc behavior in the
two-dimensional regime is still quite puzzling. From Fig.~\ref{fig:anisotropy} the
anisotropy is approximately temperature independent above $150$~K and
roughly equal to $2$. Although the constant anisotropy is indeed
to be expected if the system is in the two-dimensional regime the
{\it value} is quite surprising since it is much {\it lower} than
the ratio that would be expected from the hopping integral, and
that would be quite similar to the one actually measured for the
case of \pf6. The reason for such a low value remains to be
understood. The increase of the anisotropy (of about a
factor of two until the transition) is reminiscent of the one
occurring in \pf6\ but on a much broader temperature range
($100$~K instead of less than $50$~K for the later). Note also
that the exponents for the dc transport are quite different than
the ones for \pf6. Although \pf6\ was having the exponents
expected for two-dimensional planes in a Fermi-liquid states
($T^2$ for $a$ and $b^\prime$ axes and $T$ for the incoherent hopping
along $c^*$ direction) one finds, as shown in Fig.~\ref{fig:dcclo4V},
exponents for the three
axis between $1.15$ and $1.47$.

This difference between \pf6 and \clo4, and specially the low
value of the anisotropy ration is quite puzzling. In particular it
is expected that a universal phase diagram would hold for the
organics \cite{Jerome82}, in which one could go from one chemical
compound to the other by applying pressure. This property clearly
holds for the various instabilities. It would thus be interesting
to make a detailed comparison between \pf6 under pressure and
\clo4 as far as their transport properties are concerned to check
how much of the transport properties of \clo4 can be recovered.
Note in particular that it would be unlikely given the amount of
pressure to apply to \pf6 to change the hopping integrals
sufficiently to explain, simply by a modification of the hopping
integrals, the low anisotropy ratio observed in \clo4.

\section{Conclusion}

The comparison of the power-laws found in the temperature
dependent dc and microwave resistivity of \pf6\ and \clo4\ yields
a complex picture. The quadratic behavior found along the chains
of \pf6\ is reduced to $\rho_a(T)\propto T^{1.3}$ above the
dimensional crossover around 70 to 100~K; in \pf6\ we find a
maximum in the resistivity $\rho_{c^*}(T)$ around 80~K. The
longitudinal and transverse properties in this material are
qualitatively consistent from what it to be expected from a system
of coupled one-dimensional chains, having a crossover from a
Luttinger-liquid at high temperature to a two-dimensional (Fermi
liquid) like behavior at low temperatures.

On the contrary, even if optical data shows in \clo4\ a
one dimensional Luttinger-liquid behavior at high energy, the dc
transport shows clearly that below ambient temperature, \clo4\ is
in a two-dimensional regime. The third direction of \clo4\ is
metallic in the entire temperature range with some indications to
a semiconducting behavior above room temperature, indicating that
the crossover temperature should be above ambient temperature.
Quite surprisingly the anisotropy ratio, of about $2$, is quite
low and much lower that what would be expected from the ration of
the transfer integrals. Given the idea of a universal phase
diagram where change of chemistry would be equivalent to pressure,
these measurements suggest a careful comparison between \pf6\
under pressure and \clo4 to determine the similarities and
differences of the two systems.

\section*{Acknowledgment}
We thank G. Untereiner for the sample growth and preparation. The
$^3$He experiments have been performed by J. Thoms. The work was
supported by the Deutsche Forschungsgemeinschaft (DFG) and by the
Swiss National Fund under MANEP and Division II.

%\bibliography{totphys,zero}

\begin{thebibliography}{99}
\bibitem{remark1}Note, that although the value of 25{\%} is reported in
Ref.~\cite{Moser98}, the data actually taken from the graphs of that paper show
this value to be around 10{\%}.
\end{thebibliography}

%\end{document}

\end{document}